\documentclass[twocolumn,prl,superscriptaddress,showpacs,showkeys]{revtex4}
\usepackage{subfigure}
\usepackage{color}
\usepackage{graphicx}
\usepackage{natbib}
\usepackage{epsfig}

\begin{document}

\title{Discovery  of Localized Regions of Excess 10-TeV Cosmic Rays}

\author{A.~A.~Abdo}
\affiliation{Naval Research Laboratory, Washington, DC}
\author{B.~Allen}
\affiliation{Harvard-Smithsonian Center for Astrophysics, Cambridge, MA}
\author{T.~Aune}
\affiliation{University of California, Santa Cruz, CA}
\author{D.~Berley}
\affiliation{University of Maryland, College Park, MD}
\author{E.~Blaufuss}
\affiliation{University of Maryland, College Park, MD}
\author{S.~Casanova}
\affiliation{Max Planck Institut f\"ur Kernphysik, Heidelberg, Germany}
\author{C.~Chen}
\affiliation{University of California, Irvine, CA}
\author{B.~L.~Dingus}
\affiliation{Los Alamos National Laboratory, Los Alamos, NM}
\author{R.~W.~Ellsworth}
\affiliation{George Mason University, Fairfax, VA}
\author{L.~Fleysher}
\affiliation{New York University, New York, NY}
\author{R.~Fleysher}
\affiliation{New York University, New York, NY}
\author{M.~M.~Gonzalez}
\affiliation{Instituto de Astronomia, Universidad Nacional Autonoma de Mexico, D.F., MEXICO}
\author{J.~A.~Goodman}
\affiliation{University of Maryland, College Park, MD}
\author{C.~M.~Hoffman}
\affiliation{Los Alamos National Laboratory, Los Alamos, NM}
\author{P.~H.~H\"untemeyer}
\affiliation{Los Alamos National Laboratory, Los Alamos, NM}
\author{B.~E.~Kolterman}
\affiliation{New York University, New York, NY}
\author{C.~P.~Lansdell}
\affiliation{Institute for Defense Analyses, Alexandria, VA}
\author{J.~T.~Linnemann}
\affiliation{Michigan State University, East Lansing, MI}
\author{J.~E.~McEnery}
\affiliation{NASA Goddard Space Flight Center, Greenbelt, MD}
\author{A.~I.~Mincer}
\affiliation{New York University, New York, NY}
\author{P.~Nemethy}
\affiliation{New York University, New York, NY}
\author{D.~Noyes}
\affiliation{University of Maryland, College Park, MD}
\author{J.~Pretz}
\affiliation{Los Alamos National Laboratory, Los Alamos, NM}
\author{J.~M.~Ryan}
\affiliation{University of New Hampshire, Durham, NH}
\author{P.~M.~Saz~Parkinson}
\affiliation{University of California, Santa Cruz, CA}
\author{A.~Shoup}
\affiliation{Ohio State University, Lima, OH}
\author{G.~Sinnis}
\affiliation{Los Alamos National Laboratory, Los Alamos, NM}
\author{A.~J.~Smith}
\affiliation{University of Maryland, College Park, MD}
\author{G.~W.~Sullivan}
\affiliation{University of Maryland, College Park, MD}
\author{V.~Vasileiou}
\affiliation{University of Maryland, College Park, MD}
\author{G.~P.~Walker}
\affiliation{National Security Technologies, Las Vegas, NV}
\author{D.~A.~Williams}
\affiliation{University of California, Santa Cruz, CA}
\author{G.~B.~Yodh}
\affiliation{University of California, Irvine, CA}

\begin{abstract}
The 7 year data set of the  Milagro TeV observatory contains $2.2\times10^{11}$
 events of which most are due to hadronic cosmic rays.  This data is searched 
for evidence of intermediate scale structure. Excess emission on angular 
scales of $\sim10^\circ$ has been found in two localized regions of unknown 
origin with greater than $12\sigma$ significance.  Both regions are 
inconsistent with pure gamma-ray emission with high confidence.  
One of the regions 
has a different energy spectrum than the isotropic cosmic-ray flux at a level 
of $4.6\sigma$, and it is consistent with hard spectrum protons with an 
exponential cutoff, with the most significant excess at $\sim10$ TeV.  
Potential causes of these excesses are explored, but no compelling 
explanations are found.
\end{abstract}


\maketitle
The flux of charged cosmic rays at TeV energies is known to be nearly 
isotropic.  This is due to Galactic magnetic fields, which randomize the 
directions of charged particles.  However, numerous experiments across a wide 
range of energies have found anisotropy on large angular scales, typically 
with a fractional amplitude of $\sim10^{-3}$ (see 
\cite{Nagashima, Hall, Tibet, Macro, Kamiokande}, for example).  
Large-scale anisotropy is also seen in data from the Milagro detector 
\cite{Brian}, here we present the results of an analysis sensitive to 
intermediate angular scales ($\sim10^\circ$).

Milagro \cite{x2paper} is a water Cherenkov air shower detector located in 
New Mexico, USA at an altitude of 2630m and at $36^\circ$ N latitude.  It is 
composed of a central 60m x 80m pond surrounded by a sparse 200m x 200m array 
of 175 ``outrigger'' water tanks.  The pond is instrumented with 723 
photomultiplier tubes (PMTs) in two layers.  The top layer and outrigger 
tanks are used to determine the direction and energy, while the bottom layer 
is used to distinguish between gamma-ray induced and hadron induced air 
showers.  The outriggers, with each tank containing a single PMT, improve 
the angular and energy resolution of the detector for events collected after 
May, 2003.  Milagro has a $\sim$2 sr field of view, operates with a $>$90$\%$ 
duty cycle, and has a trigger rate from cosmic rays of $\sim$1700 Hz, making 
it well-suited to searching for anisotropy in the arrival directions of TeV 
cosmic rays.

For studies on small to intermediate scales ($\leq10^\circ$), an adaptation 
of the gamma-ray point source analysis, which has been published previously 
\cite{x2paper}, is used.  The primary difference between the previous analysis 
and the current analysis is that no cosmic-ray background rejection cuts are 
made.  These cuts removed over 90\% of the events, so the analysis reported 
here uses nearly 10 times the number of events of the previous analysis.  
Like the previous analysis, a signal map is made based on the arrival 
direction of each event.  A background map is also created using the 
``direct integration'' technique \cite{x2paper}, in which two-hour intervals 
are used to calculate the background.  Because of this two-hour interval, the 
analysis is relatively insensitive to features larger than $\sim30^\circ$ in 
right ascension (RA); a different analysis of the Milagro data sensitive to 
larger features has been performed and is presented elsewhere \cite{Brian}.

\begin{figure*}[tb]
\begin{center}
\includegraphics[width=7in]{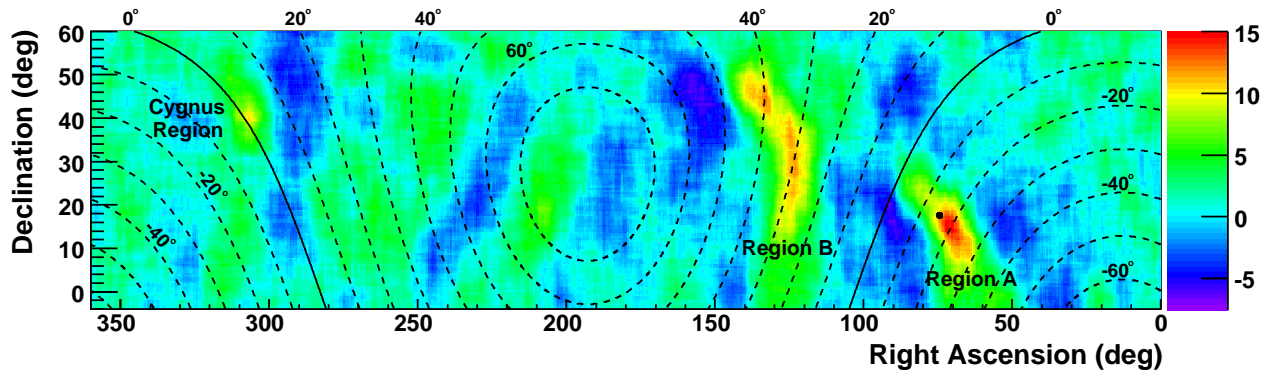}
\end{center}
\caption{Map of significances for the Milagro data set without any cuts to 
remove the hadronic cosmic-ray background.  A $10^\circ$ bin was used to 
smooth the data, and the color scale gives the significance.  The solid 
line marks the Galactic plane, and every $10^\circ$ in Galactic latitude 
are shown by the dashed lines. The black dot marks the direction of the 
heliotail, which is the direction opposite the motion of the solar system 
with respect to the local interstellar matter.  The fractional excess of 
Region A is $\sim6\times10^{-4}$, while for Region B it is 
$\sim4\times10^{-4}$.  The deep deficits bordering the regions of excess 
appear because the background calculation has been raised by the 
excess.}\label{anisky.fig}
\end{figure*}

In the gamma-ray point source analysis, the signal and background maps are 
smoothed with a square bin of size $2.1^\circ/\cos(\delta)$ in RA by 
$2.1^\circ$ in Declination ($\delta$), which is optimal for Milagro's 
angular resolution.  However, the bin size may be increased to improve the 
sensitivity to larger features, with a maximum size of about $10^\circ$ for 
$\delta<60^\circ$ (for $\delta>60^\circ$, the RA bin width 
$10^\circ/\cos(\delta)$ becomes too large for the $30^\circ$ background 
interval).  The significance is calculated using the method of Li and Ma 
\cite{LiMa}.

The analysis has been applied to data collected between July 2000 and August 
2007.  Events were required to have a zenith angle $<45^\circ$ and nFit 
$\geq20$, where nFit is the number of PMTs used in the angle fit.  With these 
cuts, the dataset consists of $2.2\times10^{11}$ events with a median energy 
of $\sim1$ TeV and an average angular resolution of $<1^\circ$.  Figure 
\ref{anisky.fig} shows the map of significances made with $10^\circ$ 
smoothing and no cuts to discriminate gamma rays from charged cosmic rays.  
The Cygnus Region, which has previously been shown to emit TeV gamma rays 
\cite{CygPaper}, is clearly visible.  The excesses labeled ``Region A'' and 
``Region B'' are seen with peak significances of $15.0\sigma$ and 
$12.7\sigma$, respectively.  These are pre-trial significances because 
the location and 
extent of the excesses were determined by examining the data. A map such as 
shown in Figure \ref{anisky.fig} has a few 100,000 independent bins, but 
given the high statistical significance many maps could be examined and the 
post trials significance would be reduced by $<1\sigma$.  The fractional 
excess relative to the cosmic-ray background is $\sim6\times10^{-4}$ for 
Region A and $\sim4\times10^{-4}$ for Region B.  Note that both excesses are 
paralleled by regions of deep deficit; this is a known effect of the analysis 
due to the fact that Regions A and B are included in the background 
calculation of neighboring areas in RA.  Therefore, the excess raises the 
background calculation above its actual value, resulting in an apparent 
deficit.

\begin{figure}[tb]
\begin{center}
\includegraphics[height=13pc]{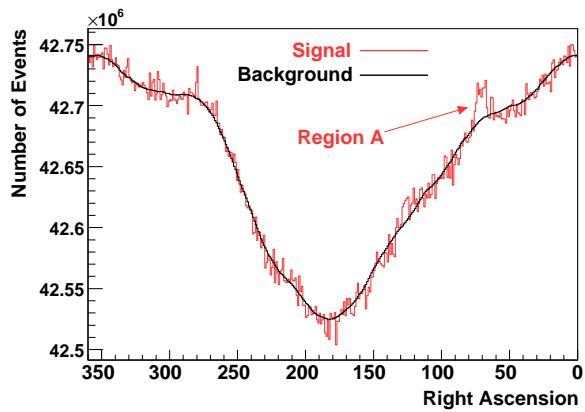}
\end{center}
\caption{Signal and background events vs RA for $10^\circ<\delta<20^\circ$.  
The plot was made using independent $10^\circ$ $\delta$ by $1^\circ$ RA bins 
(i.e. no smoothing).  A subset of the data was used in which there are only 
full days of data in order to give an approximately uniform exposure in RA.  
Region A corresponds to the excess at RA$\approx70^\circ$.  This plot shows 
that the Region A excess is inherent in the raw signal data and is not due 
to an underestimation of the background.}\label{decband.fig}
\end{figure}

Similarity is seen between the map in Figure \ref{anisky.fig} and results 
from the Tibet AS$\gamma$ collaboration \cite{Tibet,Tibet2}, but a direct 
comparison cannot be made because the analysis methods differ.  For each 
band in $\delta$, the Tibet analysis measured the excess (or deficit) 
relative to the average for that $\delta$ band, making it sensitive to the 
large-scale anisotropy discussed in \cite{Tibet}.  Smaller features, such as 
Regions A and B, were superimposed on the large-scale variation, which is 
several times greater in amplitude.  Conversely, in the analysis presented 
here, the excess/deficit was measured with respect to the local background 
calculation, which is determined from the data $\pm30^\circ$ in RA.  This is 
illustrated in Figure \ref{decband.fig}, which shows the data and background 
calculation versus RA for a $10^\circ$ band in declination without any 
smoothing applied to the data.  The large-scale variation dominates the 
figure, but the background calculation makes the analysis sensitive only to 
features with an extent smaller than $\sim30^\circ$ in RA.  It is noteworthy 
that the Tibet AS $\gamma$ collaboration has developed a model for the 
large-scale structure, and the residual map after subtracting that model 
from their data shows excesses similar to Regions A and B \cite{Tibet2}.

To estimate the extent of Region A, an elliptical Gaussian was fit to the 
excess map of the data in $0.1^\circ$ bins prior to smoothing.  The fit, which 
accounted for the change in sensitivity with declination, returned a centroid 
of RA $=69.4^\circ\pm0.7^\circ$, $\delta=13.8^\circ\pm0.7^\circ$, a half 
width of $2.6^\circ\pm0.3^\circ$, a half length of $7.6^\circ\pm1.1^\circ$, 
and an angle of $46^\circ\pm4^\circ$ with respect to the RA axis.  It is 
important to note that this fit focused on a ``hot spot'' in the general 
excess of Region A, but there is still excess extending to lower declinations.
  A fit was not performed to the excess in Region B due to its large, 
irregular shape.

While the excesses in Regions A and B are statistically significant, 
systematic causes must be ruled out.  Potential weather-related effects were 
explored by dividing the data into the four seasons, and both excesses were 
seen in each season.  The data were also divided into yearly datasets to 
investigate whether changes to the detector could play a role, and again the 
excesses were found in each dataset.  The analysis was also run using 
universal time instead of sidereal time to check for day/night effects which 
could masquerade as a signal.  In addition, the data were analyzed using 
anti-sidereal time, which provides a sanity check on the analysis since it 
will scramble real celestial features.  No excess appears in either analysis.

Potential errors in the background calculation were also investigated.  
Figure \ref{decband.fig} shows the number of events versus RA for the signal 
and background for $10^\circ<\delta<20^\circ$, using independent 
$10^\circ$ $\delta$ by $1^\circ$ RA bins (i.e. no smoothing).  The data for 
this figure were chosen to include only full days in order to achieve an 
approximately uniform exposure as a function of RA, and the broad deficit 
seen by the Tibet Air Shower Array is evident (centered around RA = 
$180^\circ$).  As can be seen, the background estimate as calculated via the 
direct integration technique \cite{x2paper} agrees well with the data.  The 
excess corresponding to Region A is clearly inherent in the raw signal data 
and is not an artifact created by the background subtraction.  A similar 
result is found for Region B.

Diagnostic tests have been performed to gain insight into the nature of 
Regions A and B.  For the purposes of these tests, Region A is defined as 
the box bounded by $66^\circ<$ RA $<76^\circ$ and $10^\circ<\delta<20^\circ$.
  Region B is defined as the union of two boxes: $117^\circ<$ RA $<131^\circ$ 
and $15^\circ<\delta<40^\circ$, and $131^\circ<$ RA $<141^\circ$ and 
$40^\circ<\delta<50^\circ$.  These definitions were chosen by visual 
inspection of Figure \ref{anisky.fig}.

To check for flux variation, the analysis was applied to yearly and seasonal 
datasets.  For each region, the yearly excess was consistent with a constant 
flux.  Both regions also had a significant excess during each of the four 
seasons, with the respective fractional excess in parts per 10000 in spring, 
summer, fall, and winter of $ 6.5\pm0.9, 4.0\pm0.9, 6.4\pm0.9, 7.2\pm0.9$ for 
region A and $3.5\pm0.4, 3.3\pm0.4, 4.0\pm0.4, ,4.7\pm0.4$ for region B. In 
both cases the fractional excess was lowest in the summer and highest in the 
winter, and the $\chi^2$ probability relative to a constant fractional excess 
is only about $5\%$ for each region.  While this may provide insight into the 
cause of these excesses, only statistical errors are given. There could be 
systematic effects such as the slightly higher energy threshold of Milagro in 
winter when there is snow on top of the pond.

\begin{figure}[tb]
\begin{center}
\includegraphics[height=28pc]{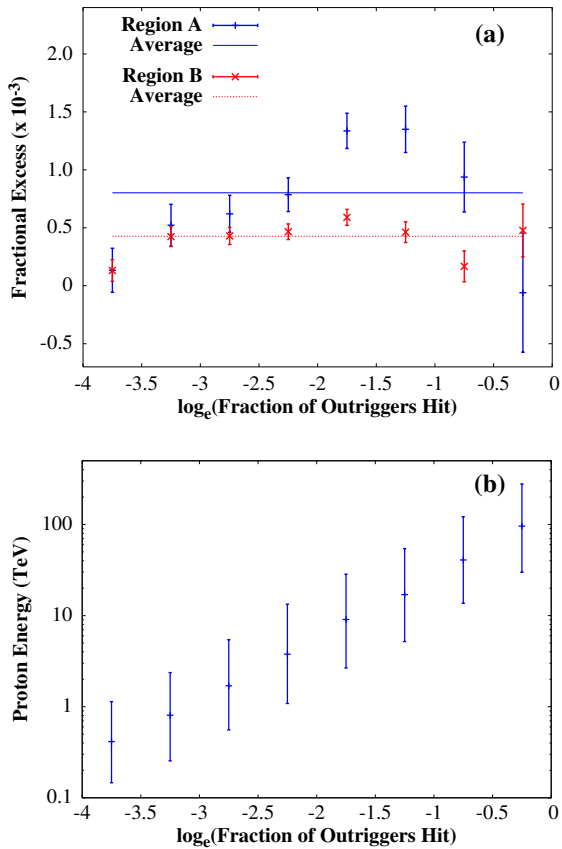}
\end{center}
\caption{(a): Differential plot of the fractional excess versus 
log$_e$(f$_{out}$) for Regions A and B, where f$_{out}$ is the 
fraction of the outriggers hit.  The spectrum of Region A is significantly 
different than the background ($2\times10^{-6}$), which is represented by the 
horizontal line. (b): Profile plot of the simulated energy of protons for the 
as a function of log$_e$(f$_{out}$). The ranges are asymmetric 
and contain the inner $68\%$ of simulated events.
}\label{logfout.fig}
\end{figure}

\begin{figure}[tb]
\begin{center}
\includegraphics[height=14pc]{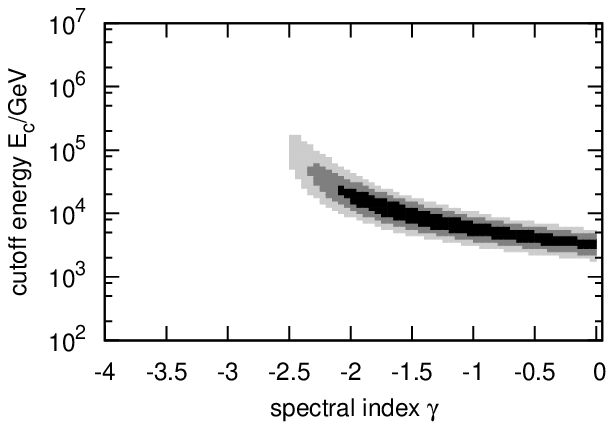}
\includegraphics[height=14pc]{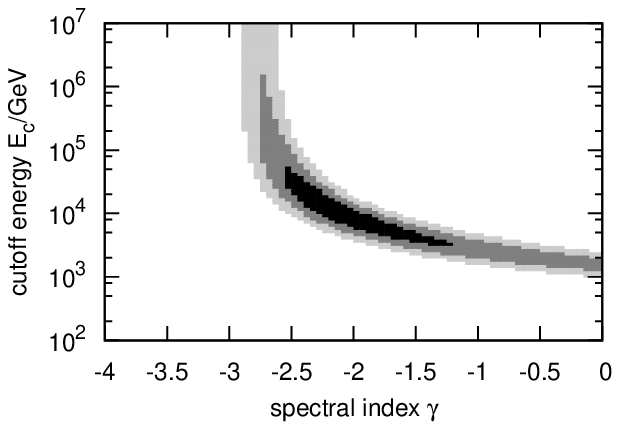}
\end{center}
\caption{Results of a $\chi^{2}$ fit to the excesses in region A and B,
assuming
a pure-proton spectrum of the form 
in Equation \ref{fitequation}.  The top panel shows the results for
region A and the bottom panel shows the results for region B.  
The $1\sigma$, $2\sigma$ and $3\sigma$
allowed regions of the spectral index $\gamma$ and the cutoff energy
$E_{c}$ are indicated by the shaded regions. 
}
\label{chi2.fig}
\end{figure}

The excesses in Regions A and B are inconsistent with pure gamma-ray emission.
We can statistically separate gamma-ray events from cosmic-ray events
utilizing two parameters.  The compactness parameter\cite{x2paper} uses
PMT information in the bottom layer of Milagro to identify the penetrating
particles characteristic of a hadronic air shower.  
The distribution of compactness depends on the energy spectrum of the source 
with higher energy gamma rays producing showers of greater compactness.
In order to exclude a gamma-ray hypothesis of any spectrum,
 we also fit an energy parameter 
f$_{out}$, the fraction of outrigger PMTs that
detect light.  Figure \ref{logfout.fig}a shows the fractional excess of
log$_e$(f$_{out}$) for regions A and B.  
We hypothesize a
spectrum for the excess of the form:
\begin{equation}
dN/dE \propto E^{\gamma}e^{-\frac{E}{E_{c}}}
\label{fitequation}
\end{equation}
where $\gamma$ is the spectral index and $E_{c}$ is the characteristic 
energy at which the spectrum cuts off.  We attempt this fit for regions A
and B assuming separately
that the primary particles are purely gamma rays and purely protons.  
The gamma-ray hypothesis in region A has a $\chi^{2}$ of 
124.0 with 16 degrees of freedom.  The cosmic-ray hypothesis
produces a reasonable fit, with a minimum  
$\chi^{2}$ of 10.3.  
For region B, the best 
gamma-ray hypothesis has a $\chi^{2}$ of 84.8 compared to a 
best cosmic-ray
hypothesis of $\chi^{2}=19.0$, again with 16 degrees of freedom.  Thus the
proton hypothesis is a reasonable fit for both regions and the gamma-ray 
hypothesis is inconsistent with probabilities of $9\times10^{-19}$ for 
region A and $2\times10^{-11}$ for region B.
The possibility that the regions contain
some admixture of protons and gamma rays has not been considered.
Figure \ref{chi2.fig} shows the $1\sigma$,$2\sigma$, and $3\sigma$ regions
around the best fit for region A and region B for the pure proton hypothesis. 
Some care must be taken in interpreting 
Figure \ref{chi2.fig}.  It does not account for our systematic errors.  
There is a estimated systematic uncertainty in the spectral 
index of $\pm0.2$ due to variation in the trigger threshold (caused 
by such things as changes in atmospheric pressure or ice on the pond).  
There is also a $\sim30\%$ systematic uncertainty in the energy scale due to 
the threshold variation, as well as discrepancy between the simulated and 
measured trigger rates.  The fit does not constrain the spectrum
well except to suggest that a hard spectrum is favored, particularly for
region A.  The cutoff energy is
constrained, with 
log$_{10}(E_{c}/GeV)$ = $4.0 ^{+0.4}_{-0.5 (stat)}$
for region A and 
log$_{10}(E_{c}/GeV)$ = $4.0 ^{+0.3}_{-0.5 (stat)}$ 
for region B.  Most importantly, the pure gamma-ray hypothesis is strongly 
disfavored.

We can see that the excesses in 
regions A and B are harder than the spectrum of the isotropic part of cosmic
rays with minimal systematic effects by looking at the data alone.
Figure \ref{logfout.fig}a shows the fractional excess in regions A and B
as a function of 
f$_{out}$.  Assuming that the excess is due to cosmic rays of the same
spectrum, we would expect the fractional excess to be completely flat.  
The offset
from zero tells us that this region is in fact an excess.  A $\chi^{2}$ test
of whether the curves in figure \ref{logfout.fig} are flat for region A(B)
returns a chance probability of $2\times10^{-6}$ 
($6\times10^{-3}$), independent of systematic errors.  
The excess of Region A is most significantly detected 
for log$_e$(f$_{out}$) $\sim-1.5$, corresponding to an energy of about 
10 TeV for protons, as shown in Figure \ref{logfout.fig}b.  At around 
10 TeV, the spectrum cuts off consistent with the results of the spectral 
fit.

There is currently no compelling explanation for the excesses in Regions A 
and B.   One possibility is that they could be due to neutrons, but this is 
unlikely because the decay length of 10 TeV neutrons is only about 0.1 
parsecs, which is much closer than the nearest star.  Another possibility is 
that these excesses could be caused by a Galactic cosmic-ray accelerator, 
but this is difficult because the gyroradius of a 10 TeV proton in a $2\mu$G 
magnetic field, which is the estimated strength of the local Galactic field 
\cite{Han}, is only $\sim0.005$ parsecs.  In order for protons from a 
cosmic-ray accelerator to reach us, the intervening magnetic field must 
connect us to the source and be coherent out to $\sim100$ parsecs since there 
are likely no sources within this distance.  However, the direction of both 
regions is nearly perpendicular to the expected Galactic magnetic field 
direction \cite{Han}  With non-standard cosmic-ray diffusion, 
it is conceivable to account for these regions with 
a nearby cosmic-ray accelerator\cite{Drury:2008ns}.

Another possibility is that one or both of the excesses could be caused by 
the heliosphere.  This explanation is supported by the coincidence of Region A 
with the direction of the heliotail 
(RA$\approx74^\circ$, $\delta\approx17^\circ$ \cite{Witte}), which is the 
direction opposite the motion of the solar system with respect to the local 
interstellar matter.  The possibility that we are seeing neutron production
in the gravitationally-focused tail of inter-stellar medium material has
been considered and discarded in \cite{Drury:2008ns} because of insufficient
target material.

In summary, Milagro has observed two unexplained regions of excess with high 
significance.  Potential systematic causes have been examined and excluded.  
Both excesses are inconsistent with pure gamma rays with high confidence, 
and their 
energy spectra are moderately to strongly inconsistent with the spectrum of 
the isotropic cosmic-ray flux.  In particular, the excess in Region A can be 
modeled as hard spectrum protons with a cutoff.  

We gratefully acknowledge Scott Delay and Michael Schneider for their 
dedicated efforts in the construction and maintenance of the Milagro 
experiment. This work has been supported by the National Science Foundation 
(under grants PHY-0245234, -0302000, -0400424, -0504201, -0601080, 
and ATM-0002744), the US Department of Energy (Office of High-Energy Physics 
and Office of Nuclear Physics), Los Alamos National Laboratory, the 
University of California, and the Institute of Geophysics and Planetary 
Physics.

\bibliography{structure}
\bibliographystyle{apsrev}

\end{document}